\documentclass[runningheads]{llncs}
\usepackage[T1]{fontenc}
\usepackage{marvosym}
\usepackage{graphicx}
\usepackage{booktabs}
\usepackage{subfigure}
\usepackage{amsmath}
\usepackage{multirow}
\usepackage[ruled]{algorithm2e}
\begin{document}
\title{Filtering with Time-frequency Analysis: An Adaptive and Lightweight Model for Sequential Recommender Systems Based on Discrete Wavelet Transform}
\titlerunning{DWTRec for Sequential Recommendation}
%
\author{Sheng Lu\inst{1,3}\orcidID{0009-0006-7222-7329} \and
Mingxi Ge\inst{1,3}\orcidID{0009-0009-4561-742X} \and
Jiuyi Zhang\inst{2,3}\orcidID{0009-0006-5620-0310} \and
Wanli Zhu\inst{1,3}\orcidID{0009-0003-0131-3244} \and
Guanjin Li\inst{1,3}\orcidID{0009-0003-6292-6730} \and
Fangming Gu\textsuperscript{\Letter,}\inst{1,4}\orcidID{0000-0001-8944-3139}}
\authorrunning{S. Lu et al.}

\institute{College of Computer Science and Technology, Jilin University, Changchun 130012 China
\\
\and
College of Software, Jilin University, Changchun 130012 China
\\
\and
\email{lusheng2122@mails.jlu.edu.cn}
\email{gemx2122@mails.jlu.edu.cn}
\email{jyzhang5523@jlu.edu.cn}
\email{zhuwl2421@mails.jlu.edu.cn}
\email{ligj2122@mails.jlu.edu.cn}
\and
\email{gufm@jlu.edu.cn}\textsuperscript{\Letter}
}

\maketitle              
\begin{abstract}
Sequential Recommender Systems (SRS) aim to model sequential behaviors of users to capture their interests which usually evolve over time. Transformer-based SRS have achieved distinguished successes recently. However, studies reveal self-attention mechanism in Transformer-based models is essentially a low-pass filter and ignores high frequency information potentially including meaningful user interest patterns. This motivates us to seek better filtering technologies for SRS, and finally we find Discrete Wavelet Transform (DWT), a famous time-frequency analysis technique from digital signal processing field, can effectively process both low-frequency and high-frequency information. We design an adaptive time-frequency filter with DWT technique, which decomposes user interests into multiple signals with different frequency and time, and can automatically learn weights of these signals. Furthermore, we develop DWTRec, a model for sequential recommendation all based on the adaptive time-frequency filter. Thanks to fast DWT technique, DWTRec has a lower time complexity and space complexity theoretically, and is Proficient in modeling long sequences. Experiments show that our model outperforms state-of-the-art baseline models in datasets with different domains, sparsity levels and average sequence lengths. Especially, our model shows great performance increase in contrast with previous models when the sequence grows longer, which demonstrates another advantage of our model.

\end{abstract}
\section{Introduction}
\begin{figure}[t]
    \centering
    \includegraphics[width=\linewidth]{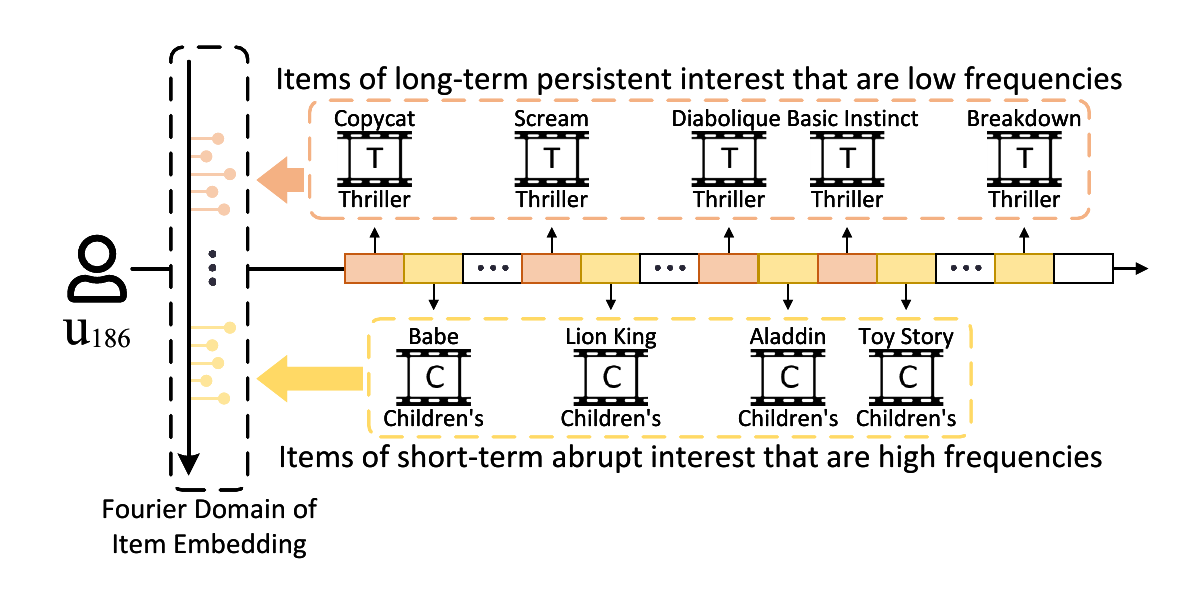}
    \caption{A user taken from ML-100k, the user's main interest is thriller films (denoted by T), but the interest sometimes switches to children's films (denoted by C).}
    \label{fig}
\end{figure}
Recommender Systems (RS) play a crucial role in our web experience \cite{surveyrecsys,boka2024survey}. Conventional RSs model user-item interactions in a static manner, which makes it challenging for them to model the dynamic user behaviors \cite{surveyrecsys,0883}. As users' interests evolve dynamically over time, it is essential for Sequential Recommender Systems (SRS) to step in. SRSs recommend by modeling the dynamic user behavior patterns behind user-item interaction sequences. In contrast with traditional models, SRSs can capture the sequential patterns of user behavior, which makes it increasingly popular among researchers and platforms \cite{alibb,sr0}.

For the advantages of SRSs, recent years have witnessed a lot of SRS proposed based on different techniques \cite{bsarec,cui2024context,gru4rec,surveygraph,s3rec}. There are models based on Markov Chains \cite{markov}, Recurrent Neural Networks 
\cite{gru4rec} or Convolutional Neural Networks \cite{caser}. With Transformer-based models showing their great performance in sequential tasks, many efforts have been put into Transformer-based SRSs \cite{sasrec,bert4rec,bsarec}. However, more recent studies reveal that the item-to-item self-attention used in Transformer is a low-pass filter in frequency domain \cite{prove1,prove2,attentionnot}, which means it only cares about the global information but lacks consideration of local characteristics \cite{fre}. In the domain of SRS, the low-pass filtering nature of Transformer makes it tend to predict items according to the users' global preferences, but in lack of consideration for users' local interests and short-term abrupt interests \cite{bsarec,fearec}. It is undemanding to capture users' long-term interests, but it remains to be challenging to predict the users' preferences by their short-term abrupt interests \cite{bsarec}. Fig.\ref{fig} shows a case study about a user in Movielens-100K dataset whose interest switches from thriller films to children's films and switches back later, which may be because of a family event or something and often occurs in other users. The sudden switches of user behavior cause abrupt changes in the sequence, from the perspective of frequency domain, the abrupt changes usually project into high frequency sub-signals, which are oversmoothed by self-attention due to its low-pass filtering nature \cite{fre,bsarec,fearec}. Further, some previous studies notice that the great deal of parameters brought by self-attention cause overfitting \cite{fmlprec}.

To tackle the low-pass filtering nature and great deal of parameters of self-attention, we look for solutions in Digital Signal Processing (DSP). We notice that some previous studies have used the Fast Fourier Transform (FFT) to address the overfitting issue and the low-pass filtering characteristic of self-attention \cite{fmlprec,fearec,bsarec}. However, FFT analyzes the
components of the signal from the whole range, and is primarily a global method
for frequency extraction rather than a localized one. As we know, a person's interests are neither kept forever nor hold same components all the time \cite{nonstation}. Traditional methods like FFT are uncompetitive in extract the components of the signal in a localized view, and are not appropriate potentially for extracting short-term abrupt interests of users. So we introduce Discrete Wavelet Transform 
\cite{waveletana,mallat} into the task of Sequential Recommendation (SR), which utilizes wavelet for localized frequency decomposition, corresponding to the need of localized interests extraction in the analysis of user-item interaction sequences.

Wavelet Transform is a widely used technique in DSP \cite{reviewwave,denoise}. Since a user-item interaction sequence are discrete, we use the discrete version of Wavelet Transform. DWT decompose a signal into sub-signals within time windows of different locations and different scales, corresponding to different durations of user interests, which is an important motivation for our designation. By DWTRec, we model user interests with the analysis of components of user-item interaction sequence, avoiding complex item-to-item computing of sequence and effectively extract the high-frequency interests of users. To adaptively filter different types of interests of users, we design an adaptive time-frequency filter which learns the weights of different interests automatically from the datasets. 

To deal with the unstable strengths of filtered signals, we design a two-step filter that reshapes the sub-signals according to time and frequency to distinguish noise and useful information for the first step, and rescale sub-signals in different hidden dimension for the second step to let the model decide the strengths of the signal in different hidden dimensions. By adaptive time-frequency filter, DWTRec can capture the important high and low frequency interests at different time. Further, DWTRec shows a prospective way to efficiently model longer sequences.

We carry out extensive studies on DWTRec to verify the efficacy of the model. We choose datasets from diverse domains with different levels of sparsity and average sequence lengths and take the State-of-the-Art (SOTA) baselines for comparison to validate the efficacy of DWTRec. We also observe great performance increase in contrast with SOTA baseline utilizing self-attention, which validates the superiority of the approach for sequential modeling of longer sequences possessing many short-term abrupt interests.

Our contributions are as follows:
\begin{itemize}
    \item To the best of our knowledge, it is the first time that a DWT-based architecture is applied in SR.
    \item We build an adaptive time-frequency filter, which is an efficient plug-and-play module to replace self-attention.
    \item We design an efficient lightweight model DWTRec for SR that utilizes time-frequency analysis to model user interests, greatly improves the SOTA.
    \item We conduct various experiments on datasets with different domains, sparsity and average lengths to demonstrate the efficacy of DWTRec. Experiments show that DWTRec outperforms the SOTA methods.
\end{itemize}
\section{Preliminaries}
\begin{table}[t]
    \centering
    \begin{tabular}{ll}
    \toprule
         Notation&   Description\\
         \toprule
         $\textit{U},\textit{V}$& 
     user and item set\\
 $\textit{u},\textit{v}$&single user and item\\
 $\textit{S}^\textit{u}$&the interaction sequence of user $\textit{u}$\\
 $\textit{d}$&number of hidden dimensions\\
 $\textit{N}$&max sequence length\\
 $\textit{Z}$&number of layers\\
 $\textit{M}$&item embedding matrix\\
 $\textit{P}$&position embedding matrix\\
 $\textit{E}^\textit{i}$&sequence embedding matrix after $\textit{i}$ layers\\
 $\textit{L},\textit{H}$&low and high pass decomposition filters\\
 $\textit{A},\textit{D}$&high and low frequency sub-signals\\
 $\textit{W}$&learnable matrices\\
 $\gamma$&decomposition level\\
 $\textit{FL}$&filter length\\
 $\textit{l},\textit{h}$&low and high pass reconstruction filters\\
 $\textit{p}(v|\textit{S}^\textit{u})$&user $\textit{u}$'s preference score on $\textit{v}$\\
 \bottomrule \end{tabular}
    \caption{Notations}
    \label{Notations}
\end{table}
To make it easy to follow the paper, we summarize the notations in Table. \ref{Notations}.
\subsection{Formalization of Sequential Recommendation}
For Sequential Recommendation, we have a user set $\textit{U}$ and an item set $\textit{V}$, where $|\textit{U}|$ denotes the number of users and $|\textit{V}|$ denotes the number of items. $\textit{u},\textit{i}$ correspondingly denote a single user or item. Along with each $\textit{u}$, there is a corresponding interaction list sorted by timestamps $\textit{S}^\textit{u}=\{\textit{v}^\textit{u}_1,\textit{v}^\textit{u}_2,...\textit{v}^\textit{u}_{|\textit{S}^\textit{u}|}\}$, where $\textit{v}^\textit{u}_1$ holds the smallest timestamp, meaning the first interacted item by $\textit{u}$. Our task is to recommend a Top-k list to $\textit{u}$ by $\textit{S}^\textit{u}$, that is to recommend items to users by their historical interactions with items. We perform this by computing and ranking $\textit{u}$'s preference score for each $\textit{v}$, which is denoted by $\textit{p}(v|\textit{S}^\textit{u})$. More clearly, we predict $\textit{v}^\textit{u}_{|\textit{S}^\textit{u}|+1}$ with $\textit{S}^\textit{u}$.
\begin{figure}
    \centering
    \subfigure[Sym6 low-pass filter]{
    \begin{minipage}[b]{0.45\linewidth}
    \centering
    \includegraphics[width=\linewidth]{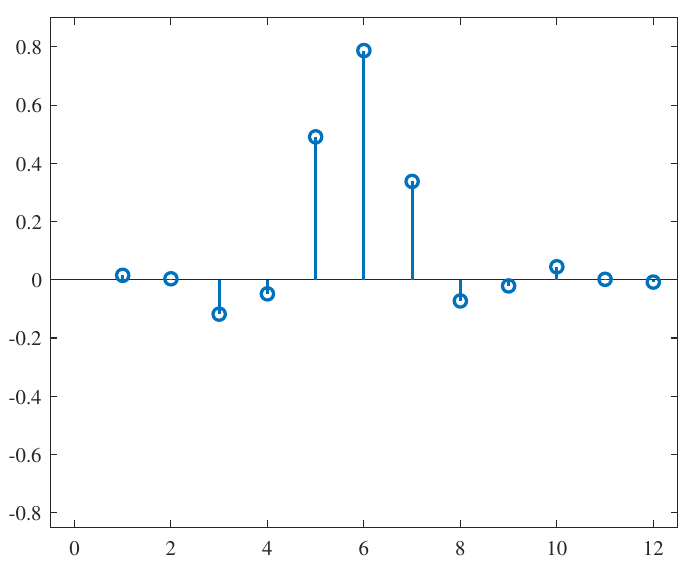}
    \end{minipage}}
    \subfigure[Sym6 high-pass filter]{
    \begin{minipage}[b]{0.45\linewidth}
    \centering
    \includegraphics[width=\linewidth]{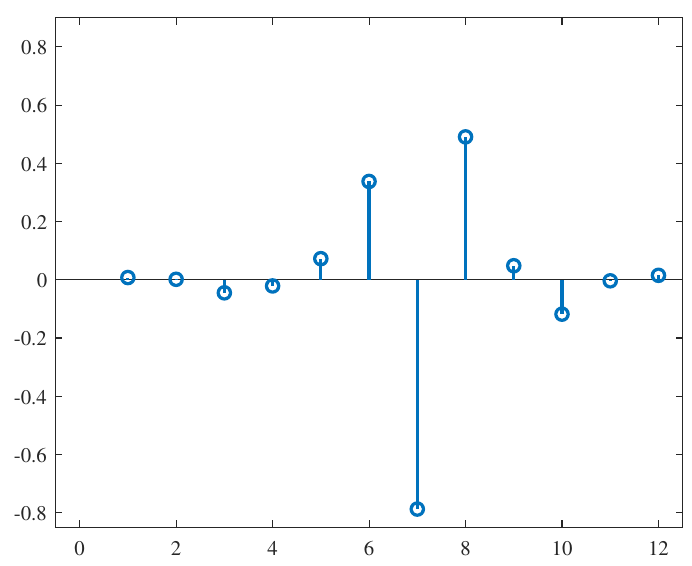}
    \end{minipage}}
    \caption{Decomposition filters of sym6 wavelet, the vertical axis is amplitude.}
    \label{wave}
\end{figure}
\subsection{The Low-Pass Filtering Nature of Self-Attention}
The core idea of self attention is to compute the weights of different $\textit{V}$s by the dot-product similarity of $\textit{Q}$ and $\textit{K}$ of corresponding items. The process is as follows:
\begin{equation}
    Attention(Q,K,V)=softmax(\frac{QK^T}{\sqrt{d}})V
\end{equation}
Recent studies reveal the self-attention is constantly a low-pass filter \cite{prove1,prove2}. In the domain of SR, the sudden switches of user interests cause abrupt changes to the signal trends, thus projecting into high-frequency information in frequency domain, which is usually smoothed out by self-attention \cite{bsarec,fearec,fre}.
\subsection{Discrete Wavelet Transform}
Discrete Wavelet Transform is the discrete version of Wavelet Transform, the main idea of DWT is to replace the infinite-length wave of Fourier Transform with a finite-length wave \cite{waveletana,meyer1992wavelets}, so the technique can extract the sub-signals within a certain period. The designed finite-length wave is called wavelet. Along with each wavelet, there is Father Scaling Function and a Mother Wavelet Function, respectively for the extraction of overall value level of the signal and detailed signal characteristics in Wavelet Transform. In DWT the two functions are replaced by a group of discrete filters including $L,l$ corresponding to the Father Scaling Function and $H,h$ corresponding to Mother Wavelet Function. The computations of i-th decomposition results are as follows:
\begin{equation}
    \textit{A}^\textit{i}[\textit{n}]=\sum^{\textit{FL}-1}_{\textit{k}=0}\textit{A}^{\textit{i}-1}[2\textit{n}-\textit{k}]\textit{L}[\textit{k}]    
\end{equation}
\begin{equation}
    \textit{D}^\textit{i}[\textit{n}]=\sum^{\textit{FL}-1}_{\textit{k}=0}\textit{A}^{\textit{i}-1}[2\textit{n}-\textit{k}]\textit{H}[\textit{k}]\\
\end{equation}
For our implementation, we use Mallat algorithm \cite{mallat} to get better efficiency. A single level decomposition is defined as follows:
\begin{equation}
    \label{eqdwt}
    A, D = DWT (\textit{X})
\end{equation}
\begin{equation}
    \label{eqdwta}
    A = (\textit{X}*L )\downarrow2
\end{equation}
\begin{equation}
    \label{eqdwtd}
    D = (\textit{X}*H )\downarrow2
\end{equation}
Here, the * denotes linear convolution, $\downarrow$ means down-sampling which is a popular technique to reduce input length \cite{downsam}. For multi-level decomposition (MWD), the process is defined as Algorithm \ref{alg:MWD}. Notice that in the algorithm, the $A$ is discarded until the last loop, as a result of according to the theory of Wavelet Transform, only $A$ of the last single-level decomposition will be used for reconstruction. After running MWD, we get $A^{\gamma},D^{\gamma},D^{\gamma-1},...D^1$. From the perspective of DSP, the frequencies of the results grow larger from left to right. $A^{\gamma}$ records the value level of the input while $D$s focus on the detailed characteristics of the input. There are two key portraits of the filter group controlling this performance that  $\sum_iL[i]^2=1$ and $\sum_iH[i]=0$. The different elements in $A,D$ mean frequency components within different time windows, the length of the time windows depend on the length of the filter and decomposition level. Fig. \ref{wave} visualizes the high and low pass decomposition filters of sym6, a popular wavelet in DSP.
\SetKwComment{Comment}{/* }{ */}
\begin{algorithm}[t]

\caption{Multi-level Wavelet Decomposition Algorithm}\label{alg:MWD}
\KwData{$X$ as input signal, $\gamma$ as max decomposition level}
\KwResult{$A^{\gamma},D^{\gamma},D^{\gamma-1},...D^1$}
$result \gets \{\}$,$i \gets 1$,$A \gets X$,$initialize \ D$\;
\While{$i \leq \gamma$}{
  \eIf{$i < \gamma$}{
    $A,D \gets DWT(A)$;\Comment{$D$ in i-th loop is $D^i$, so is $A$.}
    $result \gets result \cup \{D\}$\;
  }{
  $A,D \gets DWT(A)$\;
  $result \gets result \cup \{A,D\}$\;
  }
  $i \gets i+1$
}
\end{algorithm}
\subsection{Selection of Wavelet}
An essential composition of DWT is the filter group offered by wavelets. A lot of efforts have been put into the selection of wavelets since the proposal of the method \cite{optimumwave1,optimalwave2,reviewwave}. However, there is not a generalized rule that applies to different applications. We choose the widely used sym wavelets and coif wavelets for our implementation. For convenience and accuracy, we use pre-computed filters by \cite{pywave}.
The visualization of sym6 is shown in Fig. \ref{wave}.
As for the decomposition level, we recommend using a larger number when sequence get longer to reveal more details of the sequence.
\subsection{Comparison between Fourier Transform and Wavelet Transform}
The Discrete Fourier Transform (DFT) is an extraordinary frequency analysis method in DSP, the process of DFT for a sequence $\textit{X}$ with the length of $\textit{N}$ is as follows:
\begin{equation}
    \textit{X}_k=\sum_{\textit{n}=0}^{N-1}\textit{x}_\textit{n}\textit{e}^{-\frac{2\pi\textit{i}}{\textit{N}}\textit{nk}}
\end{equation}
Where $0\leq\textit{k}\leq\textit{N}-1$. The DFT aims to decompose a discrete sequence into different infinite waves with different frequencies. Contrastingly with DWT that decomposes inputs into combinations of finite waves with different frequencies. The DFT performs great when the components of the sequence are static, but in the domain of SR, where user interests are non-stationary \cite{nonstation}, the components of sequence are dynamic, the global analysis of DFT brings incompetent performance. Conversely, DWT gives consideration to the location of the sub-signals while telling the frequencies. As a result, DWT is our choice for building SRS.
\begin{figure}[t]
    \centering
    \includegraphics[width=0.9\linewidth]{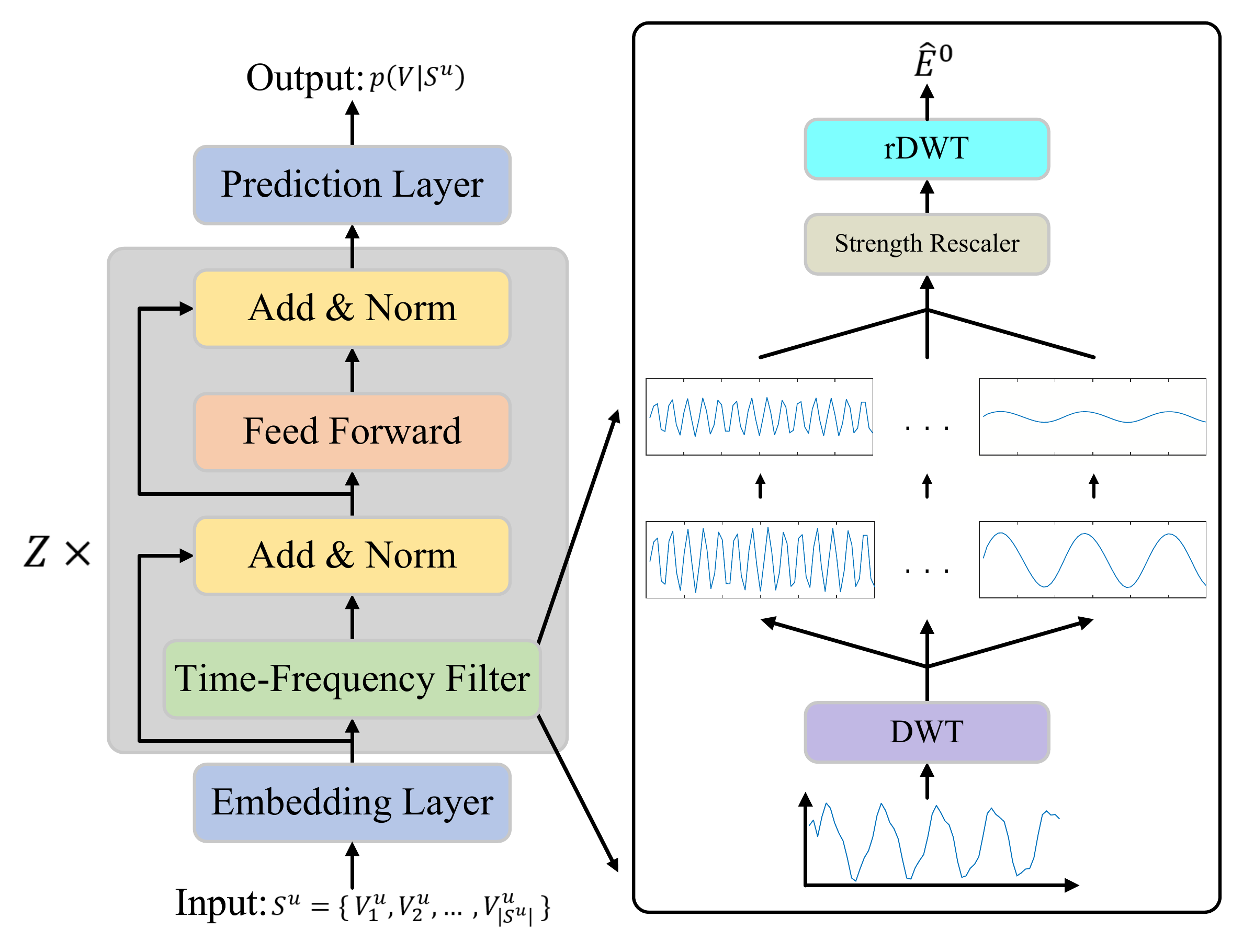}
    \caption{Overview of DWTRec}
    \label{ov}
\end{figure}
\section{Methodology}
In this section, we propose DWTRec, an efficient SR model based on time-frequency analysis, and the ideas behind DWTRec. The overview of the DWTRec method is illustrated in Fig. \ref{ov}. 
\subsection{Embedding Layer}
In the first layer of DWTRec, we aim at transforming the different inputs into sequences a fixed length, and embedding them to get a unified and comprehensive representation of each sequence for the model. Define max sequence length as $\textit{N}$. Firstly, we examine the length of input sequence $\textit{S}^\textit{u}$, if the length exceeds $\textit{N}$, then only the rightmost $\textit{N}$ items will be retained, else if the length is shorter than $\textit{N}$, we pad items with the ID of 0 on the left until the length is equal to $\textit{N}$. We define a item embedding look-up matrix $\textit{M}\in \mathcal{R}^{|\textit{V}|\times \textit{d}}$ in this layer. The embedding vector of item $\textit{v}$ is $\textit{M}_\textit{v}$. The processed $\textit{S}^\textit{u}=\{\textit{v}^\textit{u}_1,\textit{v}^\textit{u}_2,...\textit{v}^\textit{u}_{N}\}$. We look up each $\textit{v}^\textit{u}_\textit{i}$ in the look-up matrix, and concatenate them to form $\textit{E}^0\in \mathcal{R}^{\textit{N}\times \textit{d}}$. To make the values of embedding sequence position-aware, we define a learnable position encoding matrix $\textit{P}\in \mathcal{R}^{\textit{N}\times\textit{d}}$ and add it to $\textit{E}^0$. To speed up the training process and regularize the model, two widely used techniques, layer normalization and dropout \cite{ba2016layer,srivastava2014dropout}, are included in this layer. The computation is as follows:
\begin{equation}
    \textit{E}^0=Dropout(LayerNorm(\textit{E}^0+\textit{P}))
\end{equation}
\subsection{Stackable Learnable Filters Layer}
Multiple learnable filters are stacked on the embedding layer to develop a deep item encoder. A learnable filter layer is composed of two blocks, i.e., a time-frequency filter block and a feed-forward network block.
\subsubsection{Time-Frequency Filter Block}
In the time-frequency filter block, we perform filtering by decomposing the inputs into sub-signals, reshaping the sub-signals, rescaling and reconstructing them into signals with reverse DWT (rDWT). The decomposition is defined as follows:
\begin{equation}
    A^{\gamma},D^{\gamma},D^{\gamma-1},...D^1=MWD(\textit{E},\gamma)
\end{equation}
From left to right, the scales and receptive fields of results decrease, which means sequential capturing of smaller time windows, the frequencies of the results increase. Briefly speaking, left results hold more overall information about the signal, right results hold more detail information about the signal. Notice that $A^{\gamma}$ is generated by $L$, a low-pass filter, which sums up to 1 if squared, in the Wavelet Theory, that means $A^{\gamma}$ records the main structure of original signals and the information with lowest frequencies, so we avoid modifying $A^{\gamma}$ in the following processes. Inside each result, there are frequency components of different time windows and hidden dimensions. Before we define the learnable coefficients matrices, we need to know the length of each result, the length of i-th level decomposition results can be computed as follows:
\begin{equation}
length_\textit{i}=\left\{
\begin{aligned}
&N , \textit{i}=0.\\
&floor((length_{i-1}+FL-1)\div2),\textit{i}\neq0.\\
\end{aligned}
\right.
\end{equation}
Here we define the learnable coefficients matrices $\textit{W}=\{\textit{w}^1,\textit{w}^2,...\textit{w}^{\gamma}\}$, where $\textit{w}^\textit{i}\in \mathcal{R}^{length_\textit{i} \times \textit{d}}$. By the benefits of the learnable coefficients matrices, we can adaptively reshape the high-frequency sub-signals. The process is applying following computation on each $D^\textit{i}$:
\begin{equation}
    D^\textit{i}=D^\textit{i}\cdot \textit{W}^\textit{i}
\end{equation}
Where $\cdot$ notation means dot-product. By the learnable matrices, we adaptively reshape the sub-signals according to their frequencies, locations and hidden dimensions. To deal with the unstable signal strengths during the forward propagation, we design a dimensional strength re-scaler $\textit{r}\in \mathcal{R}^\textit{d}$ to let the model adjust them for itself. The process is defined as follows:
\begin{equation}
    D^\textit{i}=D^\textit{i}\cdot \textit{r}^2
\end{equation}
After filtering, the sub-signals need to be reconstructed into signals for further encoding. The multi-level rDWT algorithm is described in the Alg. \ref{rdwt}. Where $Centerkeep$ means take the central $|A^{\gamma-i}|$ elements to keep the signal length, $|A^0|$=$\textit{N}$. $\uparrow$ means up-sampling.After running this algorithm, we get $\hat{\textit{E}}=A^0$, item embedding sequences filtered by the time-frequency filter.
\begin{algorithm}[t]

    \caption{Multi-level Wavelet Reconstruction Algorithm}
    \label{rdwt}
\KwData{$A^{\gamma},D^{\gamma},D^{\gamma-1},...D^1$}
\KwResult{$A_0$ as reconstructed signal}
$i \gets 1$\;
\While{$i \leq \gamma$}{
  $A^{\gamma-i} \gets Centerkeep(l*(A^{\gamma+1-i}\uparrow2)+h*(D^{\gamma+1-i}\uparrow2))$\\
  $i \gets i+1$
}

\end{algorithm}
To prevent the gradient from vanishing when model gets deeper and get a more stable training process and better generalization ability, typical techniques i.e., skip connection, dropout and layer normalization are implemented in this layer \cite{resnet,ba2016layer,srivastava2014dropout}:
\begin{equation}
\hat{\textit{E}^\textit{i}}=LayerNorm(\textit{E}^\textit{i}+Dropout(\hat{\textit{E}^\textit{i}}))
\end{equation}
\subsubsection{Feed-Forward Network Block}
To further model the non-linear relations between the elements, we add a point-wise feed-forward network block in the stackable learnable filter layer. Gelu is a popular non-linear activation function in many sequential models these years, we apply Gelu function to add additional non-linearity. The computation is as follows:
\begin{equation}
    \label{eq:ffn}\textit{E}^{\textit{i}+1}=Gelu(\hat{\textit{E}^{\textit{i}}}\textit{W}_1+\textit{b}_1)\textit{W}_2+\textit{b}_2
\end{equation}
In Eq. \eqref{eq:ffn}, $\textit{W}_1,\textit{W}_2,\textit{b}_1,\textit{b}_2$ are all trainable parameters. Where $\textit{W}_1,\textit{W}_2\in \mathcal{R}^{\textit{d}\times \textit{d}}$, $\textit{b}_1,\textit{b}_2\in \mathcal{R}^{\textit{d}}$. Then the skip connection, layer normalization, dropout are also implemented.
\begin{equation}
    \textit{E}^{\textit{i}+1}=LayerNorm(\hat{\textit{E}^\textit{i}}+Dropout(\textit{E}^{\textit{i}+1}))
\end{equation}
\subsection{Prediction Layer}
After several layers, the item encoder has sufficiently encoded the characteristics of the input sequence. So in this final layer, we use the $\textit{E}^\textit{Z}$ to compute the preference scores for the user $\textit{u}$. The preference score of user $\textit{u}$ for item $\textit{v}$ is computed as follows:
\begin{equation}
    \textit{p}(\textit{v}|\textit{S}^\textit{u})=\textit{E}^\textit{Z}_{N}\cdot \textit{M}_\textit{v}^T
\end{equation}
We compute the preference score for each item and rank them to pick the Top-K items for recommendation. The computation is simplified with matrix computation:
\begin{equation}
    \textit{p}(\textit{S}^\textit{u})=\textit{E}^\textit{Z}_{N}\cdot \textit{M}^T
\end{equation}
\subsection{Optimization}
In this section we describe how we optimize the parameters of the model. To compute loss for optimization, we apply softmax function on the predictions $\textit{p}(\textit{S}^\textit{u})$. The criterion loss function for our model is Cross-Entropy loss. We use Adam optimizer for optimization. The loss is defined as follows:
\begin{equation}
    y_{\textit{u}}=softmax(p(\textit{S}^\textit{u})), \textit{loss}= \frac{1}{|\textit{U}|}\sum_{\textit{u}\in\textit{U}}-\log{(y_{u:g})}
\end{equation}

Where $y_{u:g}$ is the possibility of the ground-truth item, which means the label.
\section{Experiments}
To research the characteristics of DWTRec, following questions are carefully designed by us. We carry out a series of experiments to answer the questions.
\begin{itemize}
    \item \textbf{RQ1}: Does DWTRec outperform the SOTA models in diverse datasets with different domains? If so, how 
much is the improvement? (Section 4.2)
\item \textbf{RQ2}: How do key components affect the performance of DWTRec? (Section 4.3)
\item \textbf{RQ3}: Does the component-level sequential modeling improve the model's modeling ability of longer sequences in contrast with SOTA baseline? (Section 4.5)
\item \textbf{RQ4}: How is the sensitivity of hyper-parameter of DWTRec? (Section 4.4)
\item \textbf{RQ5}: How is the performance of DWTRec compared with SOTA models under the same setting? (Section 4.5)
\item \textbf{RQ6}: How is the complexity and runtime efficiency of DWTRec? (Section 4.5)
\item \textbf{RQ7}: Does DWTRec take high-frequencies sub-signals into account? Is it still learned as a low-pass filter like some of the baselines? (Section 4.5)
\end{itemize}
\begin{table*}[t]
  \centering
  \resizebox{\linewidth}{!}{
  \begin{tabular}{clcccccccccc}
    \toprule
    Datasets & Metric & Caser & GRU4Rec & SASRec & BERT4Rec & FMLPRec & DuoRec & FEARec & BSARec & DWTRec& Improv.\\
    \toprule
    \multirow{6}{*}{Beauty}&HR@5&0.0123&0.0172&0.0286&0.0472&0.0376&0.0702&0.0689&\underline{0.0711}&\textbf{0.0756}&6.3\%\\
                            &HR@10&0.0223&0.0320&0.0486&0.0727&0.0618&0.0975&0.0967&\underline{0.0982}&\textbf{0.1039}&5.8\%\\
                            &HR@20&0.0398&0.0503&0.0772&0.1047&0.0950&0.1318&0.1334&\underline{0.1337}&\textbf{0.1409}&5.4\%\\
                            &NDCG@5&0.0075&0.0118&0.0180&0.0304&0.0243&0.0507&0.0489&\underline{0.0509}&\textbf{0.0535}&5.1\%\\
                            &NDCG@10&0.0107&0.0159&0.0244&0.0386&0.0320&0.0595&0.0579&\underline{0.0596}&\textbf{0.0626}&5.0\%\\
                            &NDCG@20&0.0151&0.0205&0.0316&0.0466&0.0404&0.0679&0.0671&\underline{0.0685}&\textbf{0.0720}&5.1\%\\
    \midrule
    \multirow{6}{*}{Sports}&HR@5&0.0093&0.0117&0.0187&0.0272&0.0214&0.0395&\underline{0.0406}&0.0401&\textbf{0.0427}&5.2\%\\
                            &HR@10&0.0154&0.0180&0.0298&0.0407&0.0330&0.0558&\underline{0.0585}&0.0583&\textbf{0.0630}&7.7\%\\
                            &HR@20&0.0244&0.0321&0.0478&0.0633&0.0508&0.0781&0.0824&\underline{0.0833}&\textbf{0.0884}&6.1\%\\
                            &NDCG@5&0.0056&0.0069&0.0128&0.0179&0.0138&0.0278&\underline{0.0282}&0.0276&\textbf{0.0296}&5.0\%\\
                            &NDCG@10&0.0107&0.0092&0.0163&0.0222&0.0175&0.0327&\underline{0.0340}&0.0335&\textbf{0.0362}&6.5\%\\
                            &NDCG@20&0.0151&0.0127&0.0208&0.0279&0.0220&0.0388&\underline{0.0400}&0.0397&\textbf{0.0426}&6.5\%\\
    \midrule
    \multirow{6}{*}{Toys}&HR@5&0.0073&0.0131&0.0403&0.0438&0.0503&0.0764&\underline{0.0784}&0.0772&\textbf{0.0864}&10.2\%\\
                            &HR@10&0.0121&0.0232&0.0616&0.0642&0.0733&0.1009&0.1049&\underline{0.1058}&\textbf{0.1133}&7.1\%\\
                            &HR@20&0.0216&0.0386&0.0858&0.0946&0.1046&0.1347&0.1403&\underline{0.1432}&\textbf{0.1487}&3.8\%\\
                            &NDCG@5&0.0050&0.0080&0.0272&0.0302&0.0345&0.0569&\underline{0.0573}&0.0558&\textbf{0.0625}&9.1\%\\
                            &NDCG@10&0.0066&0.0112&0.0341&0.0367&0.0418&0.0647&\underline{0.0659}&0.0650&\textbf{0.0711}&7.9\%\\
                            &NDCG@20&0.0090&0.0151&0.0401&0.0443&0.0497&0.0732&\underline{0.0748}&0.0745&\textbf{0.0800}&7.0\%\\
    \midrule
    \multirow{6}{*}{Yelp}&HR@5&0.0099&0.0141&0.0143&0.0260&0.0164&\underline{0.0275}&0.0259&0.0267&\textbf{0.0306}&11.3\%\\
                            &HR@10&0.0178&0.0253&0.0257&0.0425&0.0287&\underline{0.0443}&\underline{0.0443}&0.0442&\textbf{0.0506}&14.2\%\\
                            &HR@20&0.0321&0.0433&0.0423&0.0689&0.0498&0.0705&0.0704&\underline{0.0713}&\textbf{0.0811}&13.7\%\\
                            &NDCG@5&0.0062&0.0086&0.0085&0.0157&0.0097&0.0169&0.0167&\underline{0.0170}&\textbf{0.0195}&14.7\%\\
                            &NDCG@10&0.0088&0.0122&0.0122&0.0215&0.0136&0.0223&0.0225&\underline{0.0226}&\textbf{0.0258}&14.2\%\\
                            &NDCG@20&0.0123&0.0157&0.0164&0.0281&0.0189&0.0291&0.0291&\underline{0.0294}&\textbf{0.0335}&13.9\%\\
    \midrule
    \multirow{6}{*}{LastFM}&HR@5&0.0284&0.0329&0.0339&0.0301&0.0385&0.0422&0.0413&\underline{0.0495}&\textbf{0.0587}&18.6\%\\
                            &HR@10&0.0376&0.0431&0.0532&0.0495&0.0633&0.0633&0.0550&\underline{0.0697}&\textbf{0.0862}&23.7\%\\
                            &HR@20&0.0550&0.0670&0.0826&0.0798&0.0853&0.0872&0.0853&\underline{0.1046}&\textbf{0.1275}&21.9\%\\
                            &NDCG@5&0.0215&0.0109&0.0243&0.0195&0.0252&0.0298&0.0273&\underline{0.0354}&\textbf{0.0421}&18.9\%\\
                            &NDCG@10&0.0245&0.0210&0.0303&0.0257&0.0330&0.0365&0.0317&\underline{0.0418}&\textbf{0.0507}&21.3\%\\
                            &NDCG@20&0.0289&0.0271&0.0377&0.0334&0.0385&0.0425&0.0392&\underline{0.0507}&\textbf{0.0611}&20.5\%\\
    \midrule
    \multirow{6}{*}{ML-1M}&HR@5&0.0912&0.1013&0.1437&0.1571&0.1343&\underline{0.2026}&0.1896&0.1871&\textbf{0.2421}&19.5\%\\
                            &HR@10&0.1629&0.1737&0.2210&0.2462&0.2166&\underline{0.2810}&0.2735&0.2781&\textbf{0.3361}&19.6\%\\
                            &HR@20&0.2483&0.2704&0.3159&0.3609&0.3325&0.3828&0.3796&\underline{0.3859}&\textbf{0.4336}&12.4\%\\
                            &NDCG@5&0.0575&0.0630&0.0949&0.0992&0.0879&\underline{0.1382}&0.1265&0.1262&\textbf{0.1687}&22.1\%\\
                            &NDCG@10&0.0806&0.0862&0.1198&0.1277&0.1143&\underline{0.1635}&0.1538&0.1556&\textbf{0.1990}&21.7\%\\
                            &NDCG@20&0.1019&0.1098&0.1436&0.1567&0.1435&\underline{0.1891}&0.1804&0.1827&\textbf{0.2237}&18.3\%\\
    \bottomrule
  \end{tabular}}
  \caption{Detailed performance of 9 methods on 6 datasets. The best results are in boldface and the second-best results are underlined. ‘Improv.’ tells the improvement against the best baseline performance.}
  \label{tab}
\end{table*}
\begin{figure}[t]
    \centering
    \subfigure[Sports]{
    \begin{minipage}[b]{0.47\linewidth}
        \includegraphics[width=\linewidth]{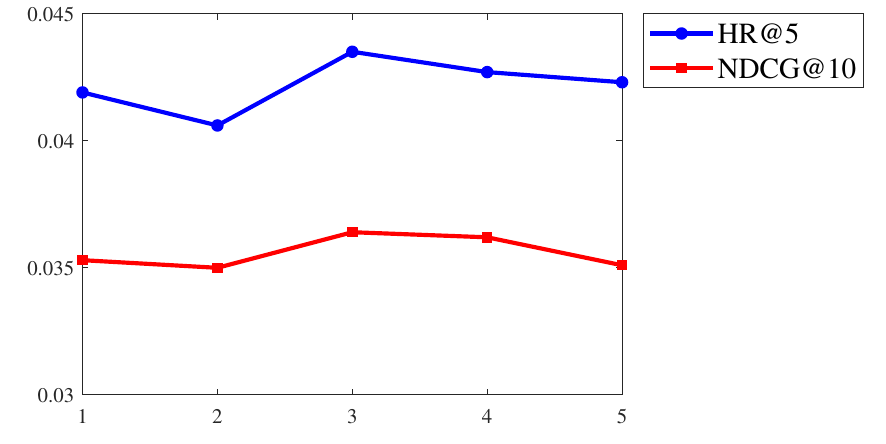}
    \end{minipage}
    }
    \subfigure[Beauty]{
    \begin{minipage}[b]{0.47\linewidth}
        \includegraphics[width=\linewidth]{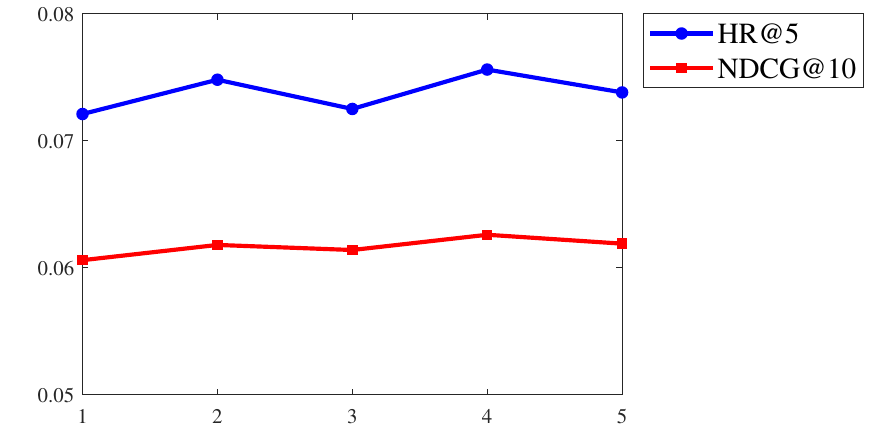}
    \end{minipage}
    }
    \caption{Visualization of performance under different $\gamma$.}
    \label{sense}
\end{figure}
\begin{figure}[t]
    \centering

        \includegraphics[width=0.6\linewidth]{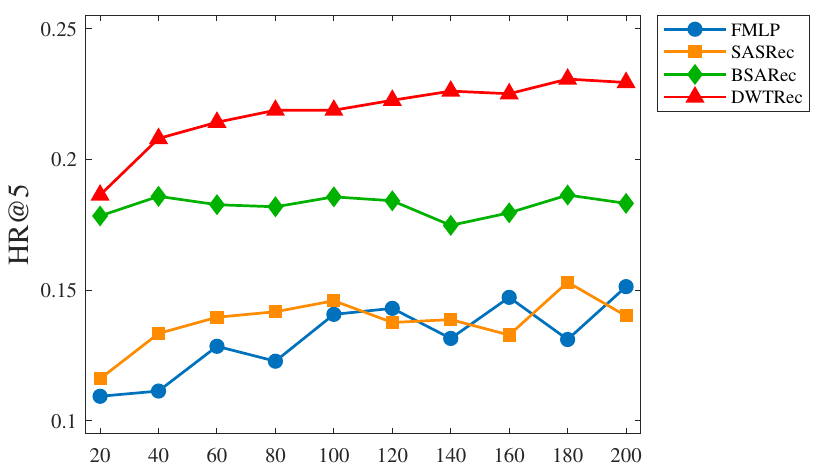}

    \caption{Performance comparison under different sequence lengths.}
    \label{sll}
\end{figure}
\subsection{Setup}
To demonstrate the efficacy of DWTRec, we carefully select SR benchmark datasets from different domains with different average sequence lengths and sparsity. Detailed dataset characteristics are offered in Tab. \ref{dataset}. To make the results convincing, we select the representative SOTA baselines of 2022 \cite{duorec}, 2023 \cite{fearec} and 2024 \cite{bsarec} and directly take their official codes for model implementation.
\subsubsection{Datasets}
Datasets are as follows: \textbf{Amazon Beauty, Sports, Toys} (e-commerce), \textbf{Yelp} (business), \textbf{LastFM} (music), \textbf{ML-1M} (movie), we follow the preprocessing steps of \cite{s3rec,fmlprec,bsarec}.

\begin{figure}[t]
    \centering
    
        \includegraphics[width=0.5\linewidth]{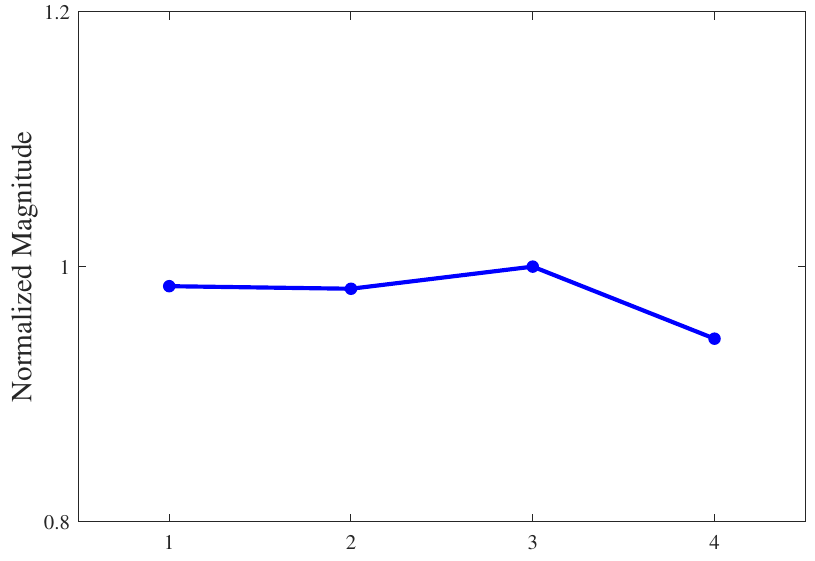}
    
    \caption{Visualization of average responses of learned filters of different levels.}
    \label{vis}
\end{figure}
\begin{table}[t]
    \centering
    \begin{tabular}{c|cccc}
    \toprule
         &   \#users&\#items&\#sparsity&\#avg.length\\
         \hline
         Beauty& 
     22,363& 12,101& 99.93\%&8.9\\
 Sports& 25,598& 18,357& 99.95\%&8.3\\
 Toys& 19,412& 11,924& 99.93\%&8.6\\
 Yelp& 30,431& 20,033& 99.95\%&10.4\\
 LastFM& 1,090& 3,646& 98.68\%&48.2\\
 ML-1M& 6,041& 3,417& 95.16\%&165.5\\
 \bottomrule
 \end{tabular}
    \caption{Detailed dataset statistics}
    \label{dataset}
\end{table}
\subsubsection{Baselines}
We select 5 kinds of baseline methods for comparison:
\begin{itemize}
    \item \textbf{CNN/RNN-based methods} i.e., Caser \cite{caser} and GRU4Rec \cite{gru4rec}.
    \item \textbf{Transformer-based methods} including SASRec \cite{sasrec} and BERT4Rec \cite{bert4rec}.
    \item \textbf{FFT-based method} including FMLPRec \cite{fmlprec}.
    \item \textbf{Transformer-based method with contrastive learning} including DuoRec \cite{duorec}.
    \item \textbf{Transformer-based methods with FFT} including two methods, FEARec \cite{fearec} and BSARec \cite{bsarec}.
\end{itemize}
\subsubsection{Metrics}
We introduce widely used SR metrics Hit Rate (HR) @K and Normalized Discounted Cumulative Gain (NDCG) @K \cite{ndcg}. The former focuses on the hit rate of the Top-K list, while the later focuses on the location of the ground truth item in the Top-K list.

\subsubsection{Implementation Details}
We carry out all the experiments on a single RTX 4090, the key packets are Pytorch 2.3.0, CUDA 12.1 and PyWavelets 1.7.0. For all the baseline models, we carefully search the optimal hyper-parameters for them based on recommended hyper-parameters. All the random seeds are set to 42 for convenience of reproduction. We promise to open-source the code, data, optimal hyper-parameters  and weights after publication for reproducibility.
\subsection{Overall Performance Comparison}

We follow \cite{duorec,fearec,bsarec} on evaluating the methods over the whole item set for fair comparison \cite{kdd}. Detailed results are in Tab. \ref{tab}.
\subsection{Ablation Studies}
To figure out how the two key components of the model work for the performance, we carry out extensive ablation studies on Beauty and Sports. By w/o Filter, we mean removing the time-frequency filter in the Learnable Filters Layer, detailed results are in Tab. \ref{tababa}. HR and NDCG are respectively denoted by H and N in Tab. \ref{tababa}. Obviously, the filter-only model outperforms FFN-only model to great extent and both modules are essential for SOTA performance.
\begin{table}[t]
    \centering
    \begin{tabular}{ccccc}
    \toprule
 & \multicolumn{2}{c}{Beauty} & \multicolumn{2}{c}{Sports}\\
 
         &  H@20&N@20 & H@20&N@20\\
         \midrule
         w/o FFN& 0.1330&
     0.0668& 0.0770&0.0376\\
 w/o Filter& 0.1162& 0.0615& 0.0704&0.0356\\
 \midrule
 \textbf{DWTRec}& \textbf{0.1409}& \textbf{0.0720}& \textbf{0.0879}&\textbf{0.0425}\\
 \bottomrule
 \end{tabular}
    \caption{Ablation Results.}
    \label{tababa}
\end{table}

\subsection{Sensitivity Studies}
In this section, the sensitivity to $\gamma$ is investigated, we set $\gamma$ from 1 to 5 at a step of 1 on Beauty and Sports and record the performance, the results are presented in Fig. \ref{sense}. Too large or too small $\gamma$ bring non-optimal results. The performance is overall stable, for applications in new datasets, a good start for search of optimal $\gamma$ is 3 or 4.
\subsection{Detailed Model Analyses}
\subsubsection{Studies On Sequence Length}
To find out whether our model outperforms Transformer-based models in modeling long sequences, we carry out experiments by setting $\textit{N}$ from 20 to 200 at an interval of 20. We choose three recent models, BSARec, SASRec and FMLP, as baseline for comparison. Results are shown in Fig. \ref{sll} which show DWTRec consistently surpasses baseline methods across various sequence lengths. We also observe a clear performance bottle neck or even degeneration on Transformer-based BSARec and SASRec when the sequence gets longer. DWTRec keeps a stable performance increment because of its ability to adaptively capture the important localized information, and provides a solution of modeling long user-item interaction sequences in sparse environments.
\subsubsection{Studies On Same Setting of Hyper-parameters}
\begin{table}[t]
    \centering
    \resizebox{!}{!}{
    \begin{tabular}{c|cccccc}
    \toprule
         &   Beauty & Sports &Toys&Yelp &LastFM &ML-1M\\
         \hline
         BSARec& 
     0.0711& 0.0401& 0.0772& 0.0267& 0.0495&0.1871\\
 DWTRec& \textbf{0.0728}& \textbf{0.0422}& \textbf{0.0796}& \textbf{0.0281}& \textbf{0.0587}&\textbf{0.2179}\\
 \bottomrule
 \end{tabular}}
    \caption{HR@5 results comparison under same hyper-parameters.}
    \label{sca}
\end{table}
\begin{table}[t]
    \centering
    
    \begin{tabular}{c|cc}
    \toprule
 & Beauty&ML-1M\\
 \hline
         DWTRec&   \textbf{5,248}&\textbf{5,248}\\
         \midrule
         BSARec& 
     16,960&16,960\\
 SASRec& 16,748&16,768\\
 \bottomrule
 \end{tabular}
    \caption{Numbers of parameter for the filters of models per layer.}
    \label{tab:my_label}
    
\end{table}
To make cogent validation about the efficacy of DWTRec, we further check the performance comparison under same model setting with BSARec, which means $\textit{N}=50, \textit{d}=64, \textit{Z}=2$. It is clear that DWTRec still makes the SOTA performance under same setting while using less parameters, and is able to resist overfitting to make better performance under larger scales. The results are shown in Tab. \ref{sca} and Tab. \ref{tab:my_label}.
\subsubsection{Model Complexity Analysis}
In this section, we analyze the complexity of DWTRec. The complexity of parameters each layer is $O(\textit{N}\gamma \textit{d}+\textit{d}^2)$. The complexity of extra space needed for runtime is $O(\textit{N}\gamma \textit{d}+\textit{N}\textit{d})$ per layer, where $\textit{N}\gamma \textit{d}$ comes from time-frequency filter and $\textit{N}\textit{d}$ comes from FFN. Contrastingly, the runtime time complexity of self-attention is $O(\textit{N}^2\textit{d})$, while the space complexity is $O(\textit{N}^2)$.
\subsubsection{Spectrum Visualization of Learned Filters}
Fig. \ref{vis} visualizes the average spectrum of filters of DWTRec on Beauty, the frequencies decreases from left to right while the magnitudes are kept. It is clear that both sub-signals of higher and lower frequencies are taken into account, indicating that DWTRec is not a low-pass filter that only cares about global information but a full-frequency filter that takes advantage of both low-frequency overall information but also high-frequency details.

\section{Conclusion}
The paper dives deep into the low-pass filtering nature of previous models in frequency domain and investigates the different types of interests of users. To capture the short-term abrupt interests of users which projects into high-frequency information while keeping low-frequency information, DWT is brought in to extract information with different frequency and time. By selectively filtering the high-frequency components of the sequence, DWTRec takes advantages of high-frequency information of different time steps. Additionally, by localized filtering, DWTRec captures different localized information of the sequence, making it perform even better on longer sequences. Extensive experiments have indeed demonstrated the high efficacy and low consumption of DWTRec.
\bibliographystyle{splncs04}
\bibliography{icic25}

\end{document}